# Optical control of energy-level structure of few electrons in AlGaAs/GaAs quantum dots

*Sokratis Kalliakos, Vittorio Pellegrini\*, Cesar Pascual Garcia*

NEST INFM-CNR and Scuola Normale Superiore, Piazza dei Cavalieri 7, I-56126 Pisa (Italy)

*Aron Pinczuk*

Dept. of Physics and Dept. of Appl. Phys. & Appl. Math., Columbia University, New York, USA

*Loren N. Pfeiffer, Ken.W. West*

Bell laboratories, Alcatel-Lucent, Murray Hill, NJ, USA

\* Corresponding author: Tel: ++39 050 509414, Fax: ++39 050 509417, email: vp@sns.it

Optical control of the lateral quantum confinement and number of electrons confined in nanofabricated GaAs/AlGaAs quantum dots is achieved by illumination with a weak laser beam that is absorbed in the AlGaAs barrier. Precise tuning of energy-level structure and electron population is demonstrated by monitoring the low-lying transitions of the electrons from the lowest quantum-dot energy shells by resonant inelastic light scattering. These findings open the way to the manipulation of single electrons in these quantum dots without the need of external metallic gates.

Research on few-electron states in semiconductor quantum dots (QDs) has been boosted by impacts of electron correlation phenomena at the nanoscale and by potential applications in quantum information processing [1-3]. Remarkable advances in this field are linked to the fact that both the confinement energy and the number of carriers can be tuned by application of metallic gates [3,4]. Transport experiments in QD systems defined by metallic gates and/or etching have highlighted the atomic-like behavior of the electronic states and provided evidences for the impact of electron-electron interaction [5,6,7]. Optical techniques have been also exploited extensively to study electron interaction phenomena particularly in self-assembled QDs [8,9,10].

Recently, evidence of large interaction effects has been reported in the spectra of spin and charge inter-shell neutral excitations probed by resonant inelastic light scattering [11,12]. These studies have provided a link between energies and intensities of these excitations and the QD confinement potential and number of interacting electrons. For these optical experiments and in order to fully exploit the capabilities of QDs for opto-electronic applications, an all-optical tuning of their electronic properties would be desiderable. This requires the development of optical approaches able to modify both the confinement energies and the number of electrons similarly to what is achieved with the application of metallic gates.

Here we demonstrate that an off-resonant laser illumination with photon frequency above the AlGaAs barrier band-gap in AlGaAs/GaAs QDs realized by etching (SEM pictures shown in Fig.1) and with few electrons yields a remarkable accurate control of these two key properties by introducing photo-generated *holes* in the QD region. We prove that a continuous modification of the QD confinement energy can be obtained in this way and we offer evidence that single-electron removal out of the QDs can be also achieved due to electron-hole optical recombination. The basis of this phenomenon is similar to that leading to the reduction of the two-dimensional electron density of a modulation-doped quantum well (QW) induced by continuous illumination with photon energy larger than the QW barrier [13-15].

Demonstration of optical tuning of the QD electronic properties is obtained by investigating the spin and charge electronic excitations among the QD energy levels by resonant inelastic light scattering in the relevant regime of few-electron occupation of the QDs. The energy positions of the observed modes are found to red-shift as the off-resonant laser power intensity increases. We argue that the photo-generated holes, accumulating at the depleted physical borders of the etched QDs, reduce the effective confinement energy of the electrons in the QD. This peculiar optical effect can be considered equivalent to that obtained by applying a positive voltage to a side gate [4]. A decrease of more than 40% can be obtained with off-resonant laser power intensities of less than 1W/cm$^2$ making this process fully compatible with experiments at Helium temperature. We also show a sudden change in the relative intensities of the observed inter-shell excitation peaks that we interpret as evidence of a change in the number of confined electrons due to optical recombination of the electrons with the photo-excited holes. Remarkably a blue-shift of the electronic excitations is instead obtained by pumping electron-hole pairs directly into the QDs by a resonant excitation below the AlGaAs band-gap. We believe that the method here presented opens novel venues for the all-optical investigation of the electronic states in QDs in the regime of few-electron occupation.

Arrays (100 μm x 100 μm) of identical QDs [16] (10$^4$ replica with inter-dot spacing of 1 μm required to improve the signal to noise ratio) are fabricated by state-of-the-art electron beam lithography and inductively-coupled-plasma reactive ion etching (ICP-RIE) on a 25 nm wide, one-side modulation-doped Al$_{0.1}$Ga$_{0.9}$As/GaAs quantum well (QW). The measured low-temperature two-dimensional electron density and mobility are $n_e = 1.1 \times 10^{11}$ cm$^{-2}$ and $2.7 \times 10^6$ cm$^2$/Vs, respectively. Scanning electron microscope images of such QDs are shown in Fig. 1. The inelastic light scattering experiments were performed in a backscattering configuration ($q \leq 2 \times 10^4$ cm$^{-1}$ where $q$ is the wave-vector transferred into the lateral dimension) at T = 2 K. A tunable ring-etalon Ti:sapphire laser with a wavelength of 789 nm in resonance with the QD absorption (the onset of inter-band QD absorption is at ~ 813 nm [16]) was focused on the QD array with a 100 $\mu$m-diameter area and the scattered light was collected into a triple grating spectrometer with CCD multi-channel detection. The off-resonant excitation was provided by a continuous-wave Helium-Neon laser at 633 nm while the photo-creation of carriers in the QD with resonant excitation below the AlGaAs band-gap (~760 nm) was obtained by increasing the intensity of the same Ti:sapphire laser used for inelastic light scattering.

In the case of GaAs QDs defined by dry etching, the number of confined electrons and the effective confinement are both determined by the large depletion region associated with Fermi level pinning due to GaAs surface states [17]. Therefore, the effective sizes of the quasi-parabolic electronic confinement are much smaller than the geometrical diameters $d$ of the etched mesa structures particularly at low-electron densities as schematically shown in the left part of Fig.1. Indeed, we have recently demonstrated [11, 16] that nanofabricated QDs prepared from the same modulation-doped QW used here and with $d \sim 200$ nm contain 3-5 electrons with a confinement energy of ~ 4 meV. In these structures therefore a depletion width of the order of 80-90 nm is present and considerably affects the electronic properties. The regime with 3-5 electrons corresponds to occupation of the first two QD energy shells.

Resonant inelastic light scattering can be applied to study neutral *monopole* (with $\Delta M=0$ where M is the total angular momentum of the confined electron system) electronic excitations in QDs that contain few electrons [11, 18]. Spin (with a change in the total spin S of the system) and charge collective excitations can be probed separately by setting the linear polarizations (parallel in the case of charge excitations and perpendicular in the case of spin modes) of the incident and scattered light [9, 16]. In the single-particle picture the excitations in these two channels are degenerate and correspond to electronic transitions between the Fock-Darwin energy levels with the same parity and separated by two times the confinement energy [11.18]. In the few-electron regime of interest here, the energy levels of the QD result from the interplay between the quantum confinement energy and the strong spatial correlation among the electrons and they can be modelled by expanding them in series of Slater determinants made by Fock-Darwin states [11,12]. In addition, the large correlation effects split in energy the spin and charge modes. The energies of these collective modes can be carefully modelled using exact diagonalization methods and then used to evaluate both the confinement energy and the degree of spatial correlation [11,12].

Representative spin and charge excitation spectra from QDs with geometrical diameter $d \sim 200$ nm are shown in Fig. 2 in red and blue, respectively. Two peaks (labelled A and B) are seen in both channels with the lowest-energy ones (A) becoming the prominent feature at large Helium-Neon intensities. The spectra in the bottom panel of Fig.2 are similar to those corresponding to the case of four electrons occupation except for the absence of the triplet to singlet (TS) inter-shell mode at around 5.5 meV [11] that is peculiar to the triplet (S=1) ground state with four electrons. This suggests that the effective electron occupation in the QDs studied here is between 3 and 5 [19] taking into account the possibility that the TS mode is too broad or weak to be resolved at the temperature of T=2K. Following the inelastic light scattering selection rules in the backscattering configuration [11, 18], we assign the peaks to the monopole ($\Delta M = 0$) inter-shell spin and charge excitations with $\Delta S = +1$ (spin) and $\Delta S=0$ (charge) from the two lowest-energy occupied shell levels as shown in the right part of Fig.2.

In the following we exploit that in this few-electron regime the excitation spectra of QDs should be very sensitive on changes of electron number and confinement energy [11, 12] and we show that these intrinsic properties can be finely tuned by illumination at 633 nm. To this end we recall that the electron density of a modulation-doped QW can be reduced by continuous illumination with photon energy larger than the QW barrier [13, 15]. The mechanism is schematically illustrated in the bottom right part of Fig. 1. It has been shown that photo-excited electrons in the AlGaAs barrier contribute to a charge compensation of the ionized donors while photo-excited holes are swept in the QW region, reducing the electron density through electron-hole radiative recombination. In the case of our etched QDs, however, due to the presence of the in-plane parabolic potential the photo-excited holes injected in the GaAs region accumulate in the depleted 80/90-nm-wide region at the borders of the etched mesa as schematically illustrated in Fig.1 (photo-generated holes are shown as empty circles). Although a quantitative analysis of this process is rather complicate given the three-dimensional nature of the confinement, we expect that the Coulomb attraction exerted by the holes to the confined electrons flattens the parabolic potential influencing the QD properties similarly to the effect of a positive voltage applied to a side gate.

Indeed Fig. 2 demonstrates that a significant reduction of the energy scale of the QD electronic states in the conduction band can be achieved with reasonable low values of the Helium-Neon power intensity. This reduction is seen in the energies of spin and charge inter-shell monopole excitations shown in Fig.2 for different Helium-Neon power densities. We recall that these inelastic light scattering measurements are performed using the Ti:sapphire laser (at 789 nm with an incident intensity of 4 W/cm$^2$) that is in resonance with our QD system and below the $Al_{0.1}Ga_{0.9}As$ bandgap. The charge excitations are blue-shifted with respect to the spin modes, due to exchange and correlation effects that are particularly relevant in the few-electron regime. Increasing the laser intensity at 633 nm, the charge and spin excitations exhibit a similar behaviour, i.e. an overall red-shift and a change of the relative peak intensities while the absolute intensity of the modes decreases due to reduction of the number of confined electrons (see Fig.3) and a possible increase in temperature induced by light. The energy splitting between charge and spin excitations does not seem to change drastically with increasing the light intensity at 633 nm although the exact determination of the energy positions of the charge peaks is hindered by the impact of the laser stray light.

The relative change of the intensities of the A and B peaks in both the spin and charge spectra is also remarkable. As the Helium-Neon intensity increases peak B in both charge and spin channels tends to disappear and all the oscillator strength for the raman-active transitions collapses on the lowest energy A mode. This is significant since the number of spin (and charge) excitations in QDs with few electrons depends on the number of electrons and only one monopole inter-shell excitation should remain in case of QDs with just two electrons and one shell occupied [11,12]. Therefore the behavior shown in Fig.2 suggests that recombination of electrons with the photo-generated holes occurs at sufficiently high Helium-Neon intensities [20] leading to the depletion of the second electronic level and bringing the QD towards the condition of population of the first shell with two electrons only. Since the QDs of the arrays might have slightly different properties (for instance a distribution in confinement energies), transitions associated to the removal of single electrons could not appear abrupt in our spectra. The contour plot of the spin excitations spectra for different Helium-Neon intensities shown in Fig.3 suggests, however, that the transition is sharp enough that single-electron accuracy can be reached by this method. This is not surprising since both the relatively sharp linewidths of the collective modes (of the order of 1 meV) and the results of the micro-photoluminescence single-dot experiments [16] indicate a large dot-to-dot homogeneity achieved in the fabrication of these QDs.

Finally, Fig.4 demonstrates the possibility of tuning confinement energy and number of electrons by adjusting the relative intensities of both the resonant Ti:sapphire at 789 nm (also used for the inelastic light scattering process) and the non-resonant laser at 633 nm. In particular Fig. 4a reports the evolution of the spin excitations as a function of the 633 nm laser intensity with a fixed 789 nm laser intensity of 50 W/cm$^2$. Comparison of these data with those shown in Fig.2 taken with 789 nm laser intensity of 4

W/cm$^2$ highlights the competition between the photo-generation mechanisms at these two different wavelengths. It can be seen that energy shifts of more than 2 meV (similar to the ones in Fig. 2a) are induced for 633 nm power densities almost one order of magnitude larger than the ones in Fig. 2a. Finally, Fig.4b demonstrates that the opposite behaviour (blue-shift) of the monopole excitations can be induced by keeping a constant power density at 633 nm while increasing the laser intensity at 789 nm. A more quantitative understanding of the effects reported here is necessary in order to optimize this approach. To this end, theoretical and experimental evaluations of the dynamics of the photo-generated electron and hole pairs in the QD in the regime of strong correlation are required.

In conclusion, we demonstrated an all-optical method for the manipulation of the lateral confinement potential and the electron number in etched modulation-doped GaAs/AlGaAs QDs with few electrons by continuous-wave illumination with photon energy larger than the AlGaAs barrier bandgap. All-optical tuning of the electronic states in these QDs could open new venues towards the understanding of few-particle effects in the exotic limits of strong interaction and weak confinement and offer new approaches for quantum information.

ACKNOWLEDGMENT  This work was supported by MIUR-FIRB No. RBIN04EY74 and by the PRIN of the Italian Ministry of University and Research. A. P. is supported by the National Science Foundation under Grant No. DMR-0352738, by the Department of Energy under Grant No. DE-AIO2-04ER46133, and by a research grant from the W.M. Keck Foundation. We acknowledge useful discussions with M. Rontani, G. Goldoni and E. Molinari.

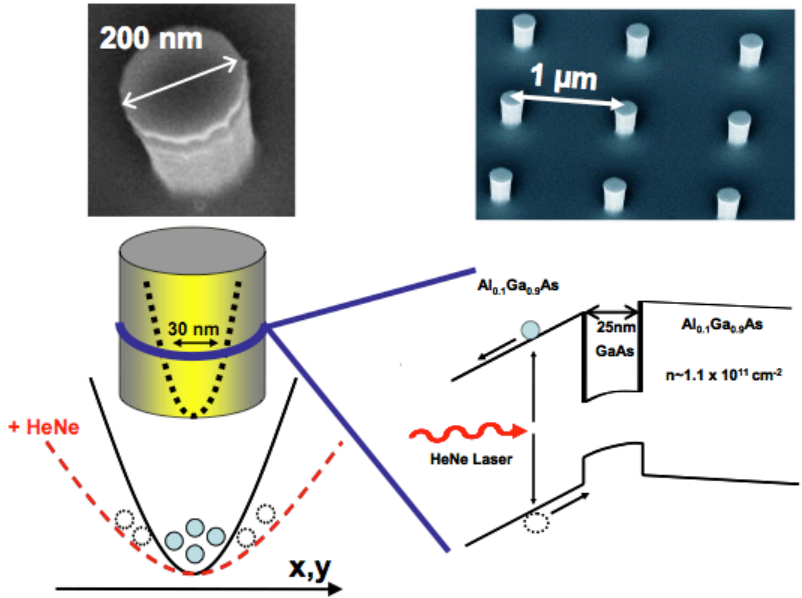

**Figure 1.** Scanning electron microscope images of AlGaAs/GaAs quantum dots (QDs) with geometrical diameter of 200 nm and inter-dot distance of 1 μm. On the bottom left a cartoon of the in-plane parabolic potential of a QD containing few electrons (filled light-blue circles) is shown. The injected holes photo-generated by illumination at 633 nm accumulate in the depletion region close to the physical borders of the etched pillar provoking a flattening of the confining potential (dashed red curve). The bottom-right part shows a schematic illustration of the optical depletion process in the quantum-well case. Electron (filled light-blue circles)-hole (dashed empty circles) pairs are created by laser illumination at appropriate wavelength (for instance He-Ne laser) in the Si-doped AlGaAs barrier. The electrons are attracted from the ionized donors inducing charge compensation while holes drift into the quantum well region.

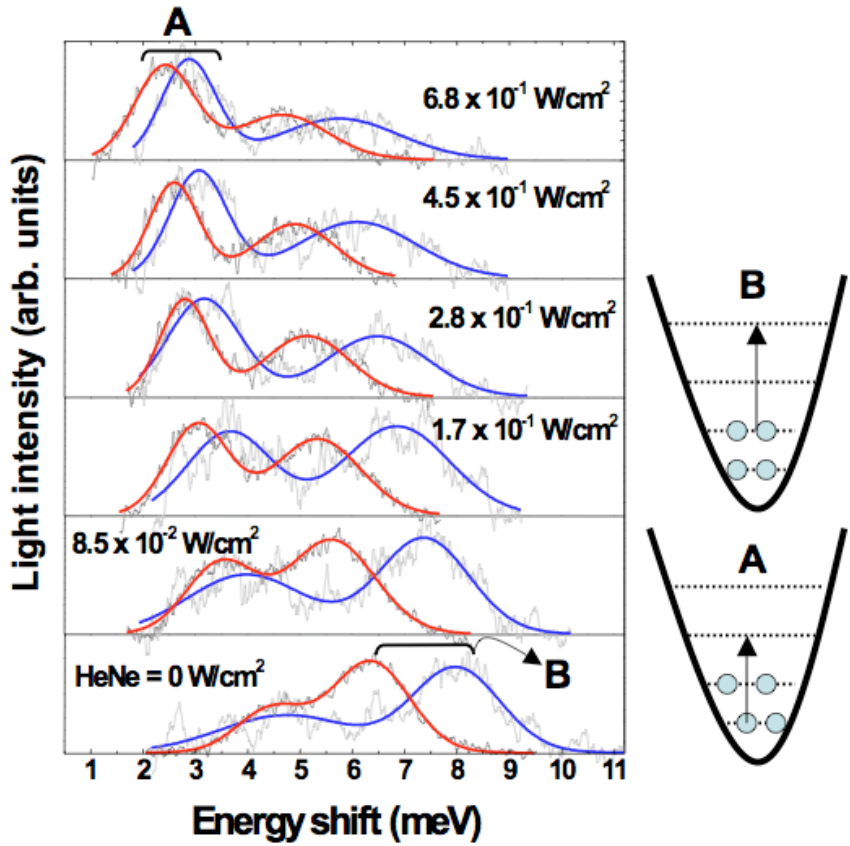

**Figure 2**. Resonant inelastic light scattering spectra at T = 2K from an array of identical QDs with geometrical diameter of 200 nm for different Helium-Neon laser intensities from zero (bottom panel) to 0.68 W/cm² (top panel). The thick lines are Gaussian fits for the spin (red) and charge (blue) excitations. The Ti:sapphire intensity used for the inelastic light scattering process was set at 4 W/cm². On the right: schematic representation of A and B raman-active excitations (vertical arrows) among the energy levels (dotted lines) with the same orbital parity (total angular momenta M) of the interacting electrons confined in the parabolic quantum dot (solid lines represent the in-plane parabolic confinement) in the case of occupation of two shell levels. Each of the two excitations is split into spin and charge modes (not shown) by exchange and correlation effects.

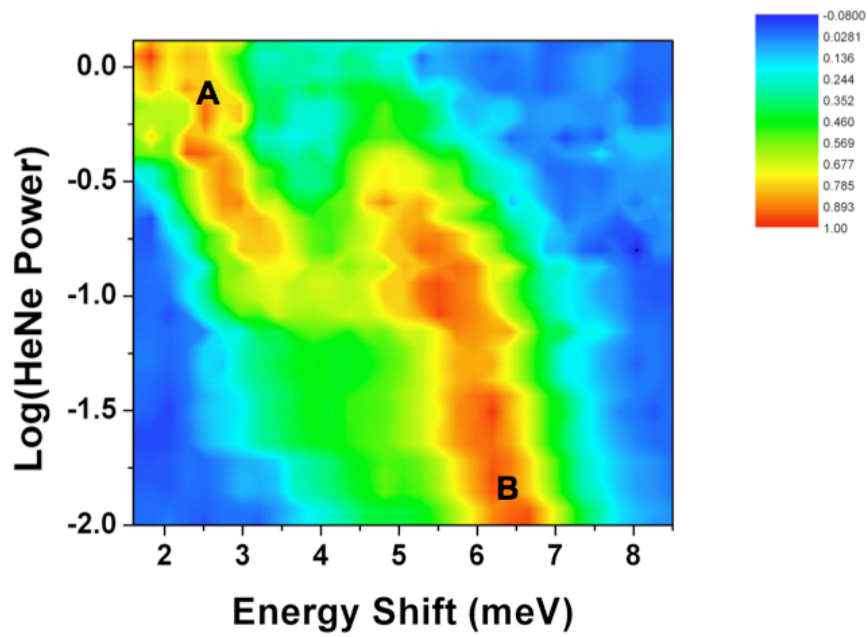

**Figure 3**. Contour plot of spin excitations A and B at T=2K for increasing He-Ne laser power (vertical axis, logarithmic scale). The Ti:sapphire intensity was set at 50 W/cm$^2$. The color indicates the light intensity ranging from blue (low intensities) up to red (high intensities).

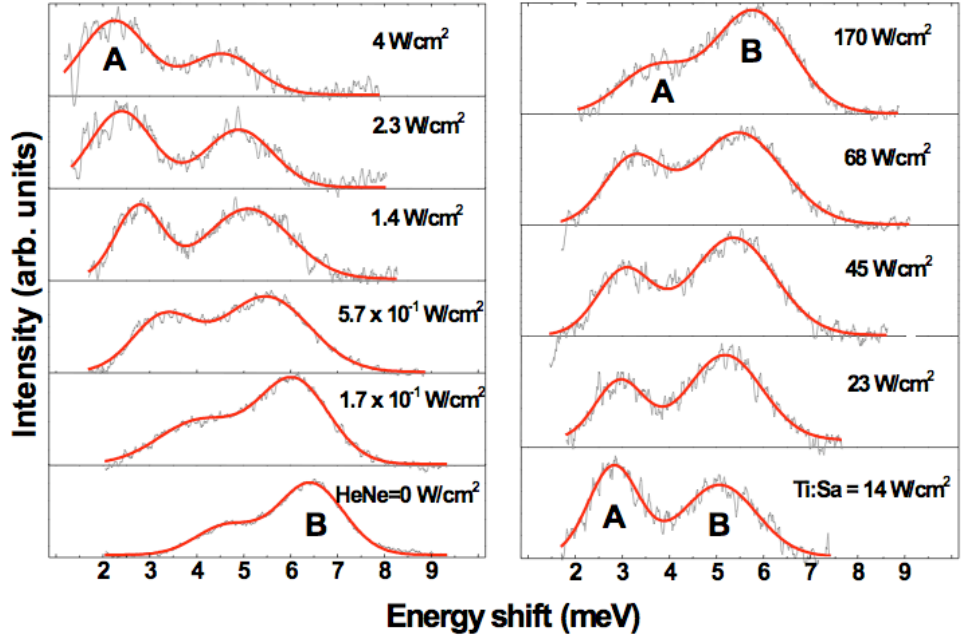

**Figure 4**. Resonant inelastic light scattering spectra of spin excitations A and B for (a) different Heilum-Neon laser intensities at a fixed Ti:sapphire intensity of 50 W/cm² and (b) different Ti:sapphire intensities at a fixed Helium-Neon intensity of 0.85 W/cm². The thick red lines are Gaussian fits. The temperature is 2K.

[19] This conclusion is supported by a detailed comparison of the spin and charge spectra shown in the bottom panels of Fig.2a with the theoretical light scattering spectra reported in Ref. [11].

[20] At Helium-Neon intensity of around I=1 W/cm$^2$ with an absorption coefficient of 2-3x10$^4$ cm$^{-1}$ and assuming that the only loss mechanism for the photo-generated holes is the exciton lifetime of around hundreds of ps we estimated an average population of less than one hole.